\documentclass[10pt,twocolumn,a4paper]{article}
\usepackage{amsmath}
\usepackage{amssymb}
\usepackage{graphicx}
\usepackage{cite}
\usepackage{color} 
\usepackage[textwidth=17.5cm,textheight=24.5cm]{geometry}

\newcommand{\vect}[1]{\boldsymbol{#1}}
\begin{document}
\title{Collective Cell Migration: Leadership, Invasion and Segregation}
\author{Alexandre J. Kabla\\\textit{Department of Engineering, University of Cambridge, Cambridge,
United Kingdom}}
\date{}

\twocolumn[
  \begin{@twocolumnfalse}
  \maketitle

\begin{abstract}
A number of biological processes, such as embryo development, cancer metastasis or wound healing, rely on cells moving in concert. The mechanisms leading to the emergence of coordinated motion remain however largely unexplored. Although biomolecular signalling is known to be involved in most occurrences of collective migration, the role of physical and mechanical interactions has only been recently investigated. In this paper, a versatile framework for cell motility is implemented in-silico in order to study the minimal requirements for the coordination of a group of epithelial cells. We find that cell motility and cell-cell mechanical interactions are sufficient to generate a broad array of behaviours commonly observed in vitro and in vivo. Cell streaming, sheet migration and susceptibility to leader cells are examples of behaviours spontaneously emerging from these simple assumptions, which might explain why collective effects are so ubiquitous in nature. This analysis provides also new insights into cancer metastasis and cell sorting, suggesting in particular that collective invasion might result from an emerging coordination in a system where single cells are mechanically unable to invade. 
\end{abstract}

\begin{center}
\end{center}  
\begin{center}
\line(1,0){470}
\end{center}
\end{@twocolumnfalse}
  ]

Many biological processes involve concerted cell displacements across large length scales and time scales. These include the most fundamental transformations taking place during embryogenesis, such as gastrulation, neurulation or vasculogenesis. Common forms of concerted motions are i) cell intercalation, through which the tissue undergoes a significant change of shape by cells exchanging neighbours (fly germ band extension \cite{Butler2009}, vertebrate gastrulation and neurulation \cite{Keller2000, Concha1998}), and ii) collective migration, during which groups of cells coordinate their direction of motion with respect to surrounding tissues, travelling reliably large distances in the embryo (neural crest cells \cite{CarmonaFontaine2008}, fish lateral line \cite{Lecaudey2008}). Collective migration also occurs in adult life, often in association with regenerative processes, such as wound healing, or pathologies, such as cancer metastasis, which often takes the form of groups of cells collectively invading other tissues \cite{Friedl2009b}. Unravelling the physical and biological principles driving collective cell migration is key to understand these critical aspects of developments, as well as trigger new therapeutic treatments for cancer. The large variability in the systems exhibiting patterns of collective migration suggests that generic principles are controlling these behaviours  \cite{Friedl2009a,Weijer2009}, and calls for an analogy with a larger class of systems \cite{Deisboeck2009}.

More generally, collective behaviours in groups of motile individuals have been widely studied over the past twenty years, both from a theoretical and experimental point of view. Systems considered include animal and human populations \cite{Couzin2003} as well as bacteria \cite{Zhang2010} or active biopolymers \cite{Schaller2010}. Flock models have shown that simple interactions between the individuals are sufficient for coordinating their direction of motion, without a need for a directional cue \cite{Vicsek1995, Gregoire2004}. One fundamental assumption in such models is the existence of a local spatial coupling, which tends to align the direction of motion of neighbour individuals. In the context of cell populations, flock models have already demonstrated that a local mechanical coupling is enough to generate streaming patterns \cite{Szabo2006, Szabo2010} and that coordination enhances the sorting dynamics of heterogeneous populations \cite{Belmonte2008}. Such emerging collective effects impose a rethink of the requirements for large scale cell coordination and in particular of how complex cell interactions must be in order to ensure robust migration in complex environments. One of the limiting factors at this stage is the difficulty to experimentally dissect the contributions of mechanical and biochemical processes.

In this paper, a numerical approach is introduced to overcome such a difficulty. A range of model studies, each highlighting a different aspect of in-vitro or in-vivo collective migration, are explored. These include in particular the transition from cohesive epithelia to mesenchymal cell populations,  the role of population size and confinement, the integration of information within the population, and the conditions for collective or solitary cell invasion in a surrounding tissue. Cell interactions will be purposefully limited to a small number of fundamental processes, such as adhesion, incompressibility of the cells, and a time scale for the evolution of cell polarity. This approach provides, as a result, a generic framework from which the role of biomolecular signalling can be reinterpreted.

\section{Results and Discussion}

\subsection{Dense endothelial cell monolayers}

The migratory patterns of cultured endothelial cell monolayers are well documented, confirming the emergence of collective behaviours such as streaming and large scale velocity correlations \cite{Haga2005, Poujade2007, Vitorino2008, Angelini2010, Angelini2011}. These experiments have shown in particular that cell density is one of the key parameters controlling collective behaviours. In vivo, cells are however usually closely packed, forming small confined groups within larger tissues \cite{Friedl2009a,Weijer2009}. The onset of group migration then primarily involves variations of motile behaviour, adhesive properties or environmental conditions \cite{Friedl2009a}, rather than cell density. The first question addressed concerns the role of these physical and mechanical parameters in the dynamics of large and dense cell populations. The effect of the population size and its mechanical environment will be subsequently studied. 

The numerical tissue is made of a 2D confluent layer of cells on a substrate. Cells can adhere to each other, and the resulting cohesion of the tissue is accounted for by a membrane tension, $J$, which is homogeneous across the population. $J$ controls here a number of otherwise independent properties, such as the adhesion energy, the cortical tension, and the amplitudes of membrane fluctuations \cite{Krieg2008}. Moreover, all cells have the same volume and cannot overlap each other. In this first section, all cells also generate the same motile force $\mu$ on the substrate. However, each cell has its own direction of polarization $\vect{n}_i$ along which this force is produced  ($|\vect{n}_i|=1$). The population lives on a surface with periodic boundary conditions, such that if a cell leaves on one side, it reappears on the opposite. As a result, although the system has a finite size, it nevertheless has no boundary that could influence cell trajectories. The Cellular Potts Model (CPM) \cite{Graner1992}, which has already demonstrated its broad biological relevance for epithelial tissues, is used to implement these rules (see Materials and Methods). 

No specific molecular mechanism for inter-cellular communication is considered. Only excluded volume interactions and constraints of membrane geometry cause cells to feel each other or feel any other physical obstacle. It is however assumed that there is a feedback from earlier displacements to the polarization itself, so that the cell tries to move towards its direction of polarization, but the polarization itself evolves towards the cell's net displacement with a response time $\tau$ (see Materials and Methods).  $\tau$ acts as a persistence (or memory) time for the cell polarization. These assumptions, first introduced by Szabo et al. in 2006 in a self-propelled model \cite{Szabo2006}, are consistent with a number of recent observations \cite{McCann2010, VanHaastert2010}, and provide a simple way to model the interaction between a cell and its mechanical environment. Szabo et al. \cite{Szabo2010} have used such a model to study the dynamics of endothelial cell monolayers. They have in particular explored the role played by the persistence time in the resulting streaming behaviour and successfully replicated the experimental shape of the velocity correlations in homogeneous cell populations using realistic values of the model parameters. This demonstrates the general potential of the self-propelled cellular models for the study of collective migration patterns. 

\begin{figure}[!ht]
\begin{center}
\includegraphics[width=3.4in]{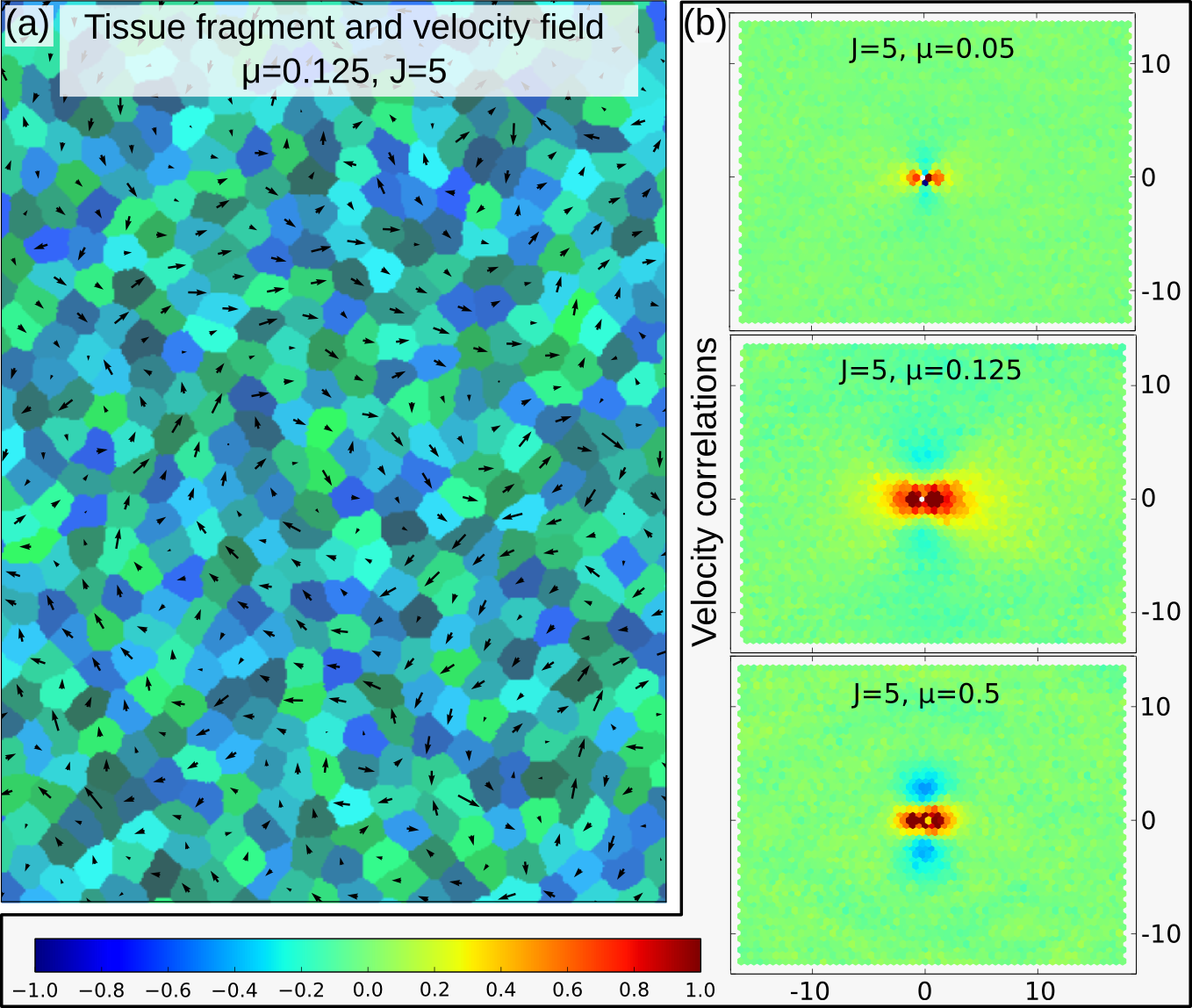}
\end{center}
\caption{(a) Image of a motile tissue in the steady state, overlayed with the corresponding velocity field. Cell colors are arbitrary. (b) Maps of the velocity correlations around a cell migrating towards the right (horizontal axis goes from back to front). The unit distance corresponds to the cell diameter. \label{velocities} }
\end{figure}

The role of the motile force $\mu$ on the dynamics of a large cell population is first considered, with an emphasis on qualitative transitions, rather than quantitative comparisons with experimental data. In the following, the persistence time $\tau$ and noise level are kept constant. At low motile forces, no net motion is observed
; the population essentially behaves like a cohesive epithelium, with a strong persistence of the neighbourhood relationships. At higher motile forces, a transition occurs to a regime where cells migrate on the substrate and stream over large distances (see figure \ref{velocities}a and Video S2 \cite{SupMat}). The spatial correlations of the velocity field quantify the structure of the streaming patterns. Considering a cell at the origin moving towards the right, the average degree of alignment between the velocity of this cell and the velocity of the cell located at $\vect{\delta r}$ is given by the function $g(\vect{\delta r})$: $g=1$ means full alignment, $g=-1$ corresponds to cell running in opposite directions, and $g=0$ to a non-correlated situation (see Materials and Methods). Figure \ref{velocities}(b) shows 2D color-coded maps of correlation functions for three different values of the motile force $\mu$. 

For intermediate values of the motile force ($\mu \approx 0.1$), strong correlations in the velocity field develop, in particular along the direction of the cell displacement (streaming), whereas displacements tend to be anti-correlated in the normal direction (meaning that streams of opposite direction are running next to each other). Such patterns are typically observed in large cell populations \cite{Haga2005, Vitorino2008, Szabo2010} and in-silico \cite{Szabo2010}. When varying (at constant density) the motile force towards smaller or larger values, we find however that spatial correlations tend to become more localized, and could decay as fast as over a few cell diameters. The non-monotonic evolution of the correlation patterns shows that the dynamics of such a dense cell population involves several competing effects that will be developed further in the following sections.

\subsection{Emergence of collective motility}

\begin{figure}[!ht]
\begin{center}
\includegraphics[width=3.2in]{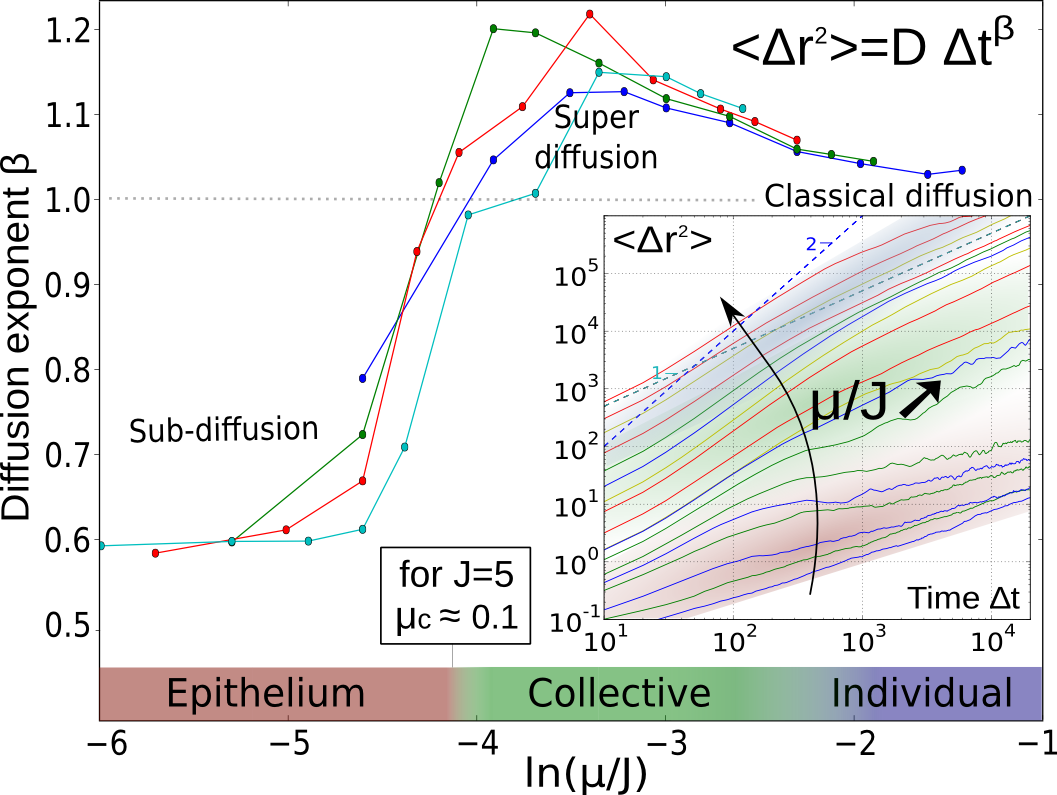}
\end{center}
\caption{Graph of the diffusion exponent $\beta$ as a function of $\mu/J$, for four different values of $J$ (2.5, 5, 7.5 and 10 - arbitrary units). Inset: average mean square displacement as a function of time for a range of motility $\mu$ and adhesion $J$. Time is expressed in MCS (see Materials and Methods) and $\tau=10$ MCS.\label{diffusion}}
\end{figure}

The transition from a non-motile to a motile population can be captured by the temporal correlations of the cell velocities, or equivalently by quantifying the evolution of the mean square displacement of the cells $\Delta r^2(\Delta t, J, \mu)$, between two time points $t$ and $t+\Delta t$ (see inset of figure \ref{diffusion}). For a given pair of parameters ($\mu$, $J$), the mean-square displacement curve is fitted over the time range $10<\Delta t<10^4$ by the function $\Delta r^2 = D \Delta t^\beta$. The value of $\beta$ characterizes the nature of cell displacements over that time interval. $\beta(\mu,J)$ is reported in figure \ref{diffusion}. For reference, $\beta=1$ indicates a stochastic, diffusive like, behaviour. The data show that for each value of $J$, there is a critical motile strength $\mu_c$, scaling with $J$, below which transport is sub-diffusive ($\beta<1$), i.e. where cells essentially never move more than their own size. For $\mu \gtrsim \mu_c$, migration is however super-diffusive, indicating that the direction of motion is persistent over large time-scales, several hundred times the memory time ($\tau=10$ in these simulations). This regime is consistent with the streaming patterns discussed in the previous paragraphs. Eventually, as $\mu$ increases further, the behaviour progressively converges towards a simple diffusive behaviour. At a qualitative level, the presence of an optimum for the collective properties reflects two competing processes taking place simultaneously: i) as the motile force increases, the coupling between cells also increases, facilitating coordination and triggering collective migration. However, ii) highly motile cells can also easily disrupt the coordination of motile groups by penetrating them. This regime of high motile forces will be explored further in the context of cell invasion (section \ref{invasection}). The typical shape of the mean-square displacement curves reported in figure \ref{diffusion} is in agreement with the theoretical predictions of Peruani and Morelli \cite{Peruani2007} for systems with uncorrelated fluctuations for cells speeds and directions of motion; speed fluctuations are here related to the random membrane fluctuations, whereas the polarization evolves over larger scales, depending on the cell environment and the emerging streaming patterns. These different scales of fluctuations are also qualitatively consistent with in vitro studies \cite{Szabo2010}, although living systems exhibit more complex signatures, in particular due to cell divisions \cite{Angelini2011}.

The emergence of streaming patterns, observed in most self-propelled models, is not in itself surprising. However, interpreted within the context of a cell population, these results show that i) local coordination can arise from purely mechanical interactions, and that ii) there is a discrete transition from a static epithelium to a streaming population. This behaviour is reminiscent of the Epithelial-Mesenchymal Transition (EMT) in living systems, in which cell behaviour changes rapidly from non-migratory to migratory. The causes of such a transition in vivo are still highly debated \cite{Tarin2005,Thiery2006}. Although a large number of oncogenes are involved, the EMT does not seem to be associated with a proper lineage switch. Figure \ref{diffusion} shows that a number of \textit{continuous} physical quantities, such as the motile force or the cohesion of the cellular layer,  control a rapid transition between two \textit{discrete} dynamical states, from a static tissue to a highly mobile population of cells, and vice-versa. It appears that, even in such a simplistic model, a significant change in a tissue phenotype can originate from minute modulations of either the cells' behaviour or their environment, without a need for a cell-lineage switch; an extrapolation to more realistic 3D situations might explain the practical difficulty in finding robust determinants for the EMT, in particular based on gene expression.

\subsection{Emerging consensus and size effects}

\begin{figure}[!ht]
\begin{center}
\includegraphics[width=3.2in]{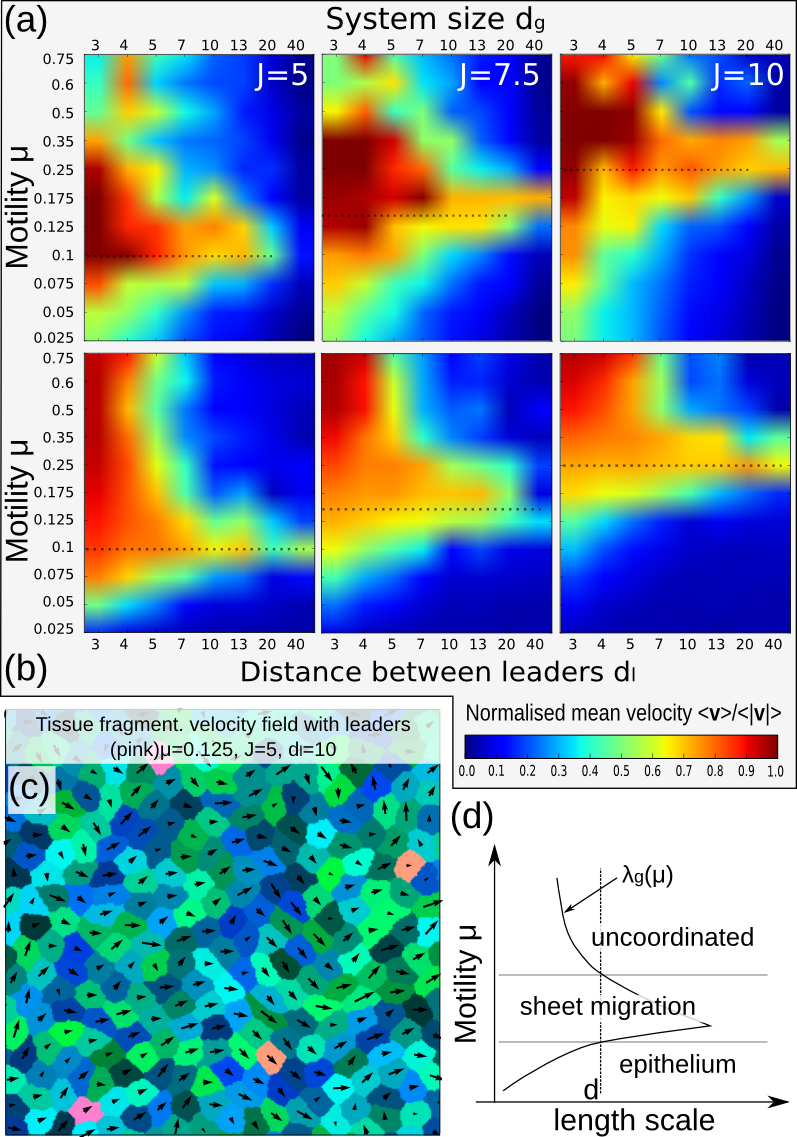} 
\caption{\label{leaders} (a) and (b): Heat maps of the order parameter $\left< \vect{v} \right> / \left< |\vect{v}| \right>$ as a function of (a) $\mu$ and the system size $d_g$ (number of cells: $N\approx d_g^2$) or (b) $\mu$ and the typical distance between leaders $d_l$ (in cell diameters). Data for three different values of $J$ is presented. (c) Example of a tissue with a few leader cells (with pink/orange tone) whose polarity is constant and directed towards the right. (d) Sketch of the curve $\lambda_g(\mu)$ and its qualitative relationship with the different regimes of migration. For a given length scale $d$ associated with a constraint (distance between leaders, distance between boundaries, or number $N$ of cells in the group ($d=\sqrt{N}$)), three regimes can be defined as $\mu$ increases: epithelium, sheet migration, or uncoordinated.}
\end{center}
\end{figure}

In vivo cases of collective migration usually involve small groups of cells with highly coordinated directions of migration, for instance during the collective invasion of carcinoma cells, or in a number of developmental processes such a neural crest cell migration or lateral line primordium (migrating epithelium). In order to probe further the effect of the population size, systems containing from 9 to 1600 cells with periodic boundary conditions have been studied. The substrate size was changed accordingly so that the cell density remained constant (see Video S3 \cite{SupMat})
. The degree of global coordination is quantified as the mean velocity across the whole population normalized by the mean cell speed ($\left<\vect{v}\right>/\left<|\vect{v}|\right>$). This corresponds to an order parameter, taking values from 0 (no order) to 1 (full coordination or sheet migration). Figure \ref{leaders}a shows how this parameter depends on the motility, for various values of $J$. For each set of parameters, one can identify a typical length scale $\lambda_g(\mu, J)$ corresponding to the largest system size at which global coordination spontaneously arises. As expected from the previous results, the ability to align is maximum when the system is at the onset of collective migration ($\mu \gtrsim \mu_c$), i.e. when temporal and spatial correlations are at their maximum. Sheet or group migration appears therefore as an intrinsic behaviour of groups of motile cells: for a small cell population, an increase of cell motility leads to a spontaneous transition from a non-motile epithelium to a migrating sheet. 
These results might shed light on recent observations of sheet migrations in vivo, for instance in the fish lateral line primordium \cite{Lecaudey2008}. The model predicts in particular that a progressive increase of the traction force (or decrease of cohesion energy $J$) in an epithelium leads first to a regime of high coordination, where sheet migration is relatively straightforward to achieve.

\subsection{Leader cells and integration of external cues}

Coordination alone is however not enough to prescribe any particular direction of motion for the population. Setting a directional preference for a group of cells requires some level of interaction with the environment, for instance to sense a gradient. In the context of collective migration and invasion, only a small proportion of competent (leader) cells in the population seems to be required in order to induce directed motion \cite{Gaggioli2007, Poujade2007, Vitorino2008}. However, how these leaders might steer the whole population remains unclear. In this section, physical mechanisms by which leader cells could influence large scale migratory patterns are explored. 

Within our framework, leader cells would be cells whose polarity is set by an external cue, rather than through a feedback from their previous displacement; their direction of polarity $\vect{n}$ is therefore kept constant, aligned along a fixed arbitrary direction, identical for all leader cells. These cells are otherwise indistinguishable from the other cells (including same motile force $\mu$ and tension $J$). By contrast with live situations where leader cells tend to be located at free boundaries, a small population of such leader cells is here scattered uniformly in the tissue (see figure \ref{leaders}c and Video S4 \cite{SupMat}). This choice allows us to study specifically how information propagates through the population; modelling a free surface would introduce many additional parameters which would make the process less tractable at a qualitative level. 

Considering large populations (too large to spontaneously coordinate), the coordination parameter $\left<\vect{v}\right>/\left<|\vect{v}|\right>$ is now calculated as a function of the cell motile force $\mu$ and the distance $d_l$ between leader cells (see figure \ref{leaders}b). This distance is now the most relevant length scale, related with the leader density $\rho_l\approx d_l^{-2}$. The larger the leader density (or the smaller the distance between leaders), the stronger is the resulting coordination. However, as expected from systems of self-propelled particles \cite{Couzin2005}, the susceptibility of the population to leader cells strongly depends on the collective aspect of the dynamics; when $\mu \approx \mu_c$, 1\% of leader cells are sufficient to significantly coordinate the whole cell population. As previously, one can define a length-scale for each set of parameters, $\lambda_l(\mu,J)$, corresponding to the largest distance between leaders that can induce a global motion of the population. A comparison between the figures \ref{leaders}a and \ref{leaders}b shows that the susceptibility to leader cells strongly correlates with the intrinsic ability of the population to coordinate their direction of motion. A population will respond well to leader cells if $d_l < \lambda_l(\mu,J) \approx\lambda_g(\mu,J)$. The population's response to leader cells is therefore primarily controlled by its own internal dynamics, encompassed by the length-scale $\lambda_g(\mu,J)$. Any constraint that acts over a length scale $d$ smaller than  $\lambda_g(\mu,J)$ is susceptible to trigger global coordination, as summarized in figure \ref{leaders}d. In particular, any persistent directional bias (leader cells, anisotropic boundaries, etc) could steer the overall migration, as long as coordination is already ensured by the collective dynamics of the group. 

Another implication of this model is that neither specific mechanisms of communication between leaders and normal cells, nor a long range mechanical coupling through the substrate \cite{Angelini2010}, are required for large scale coordination; the same short-range mechanical interactions are enough to serve that purpose. The idea that simple physical/mechanical interactions could indeed play a role in the leader/follower relationship has been mentioned \cite{Lecaudey2008, Poujade2007, Deisboeck2009}, and the results presented above demonstrate that such a scenario is indeed plausible. The behaviour observed is moreover in very good agreement with the in vitro studies of Vitorino and Meyer \cite{Vitorino2008} who have shown that i) local coordination within 2D populations of endothelial cells is independent of the presence of directed migration in the population, and that ii) only a very small proportion of leader or ``pioneer" cells are required to steer the whole population of cells. 

\subsection{Tissue invasion}
\label{invasection}
Collective forms of tissue invasion not only involve the coordination of a population of cells, but also its ability to penetrate a surrounding matrix or tissue. A number of physical and biological factors are known to influence collective invasion \cite{Friedl2009b}, but a general picture remains challenging to extract at this stage. For instance, both the intrinsic behaviour and the environment of cancer cells seem to determine the collective or individual nature of the invasion process. In this part, the framework introduced above is adapted to study cell invasion in a passive surrounding tissue. The tumour cells are essentially a population of motile cells enclosed by a tissue. The latter is modelled for simplicity as an epithelium of non-motile cells with the same mechanical properties (encompassed by $J$) as the tumour cells, so that adhesion based sorting effects will not occur \cite{Graner1992}. $J$ is kept constant in what follows; its role would simply be to shift the transitions as observed on figures \ref{diffusion} and \ref{leaders}.  The efficiency of invasion as a function of the motile strength of the tumour cells is now characterized.

\begin{figure}
\begin{center}
\includegraphics[width=3.0in]{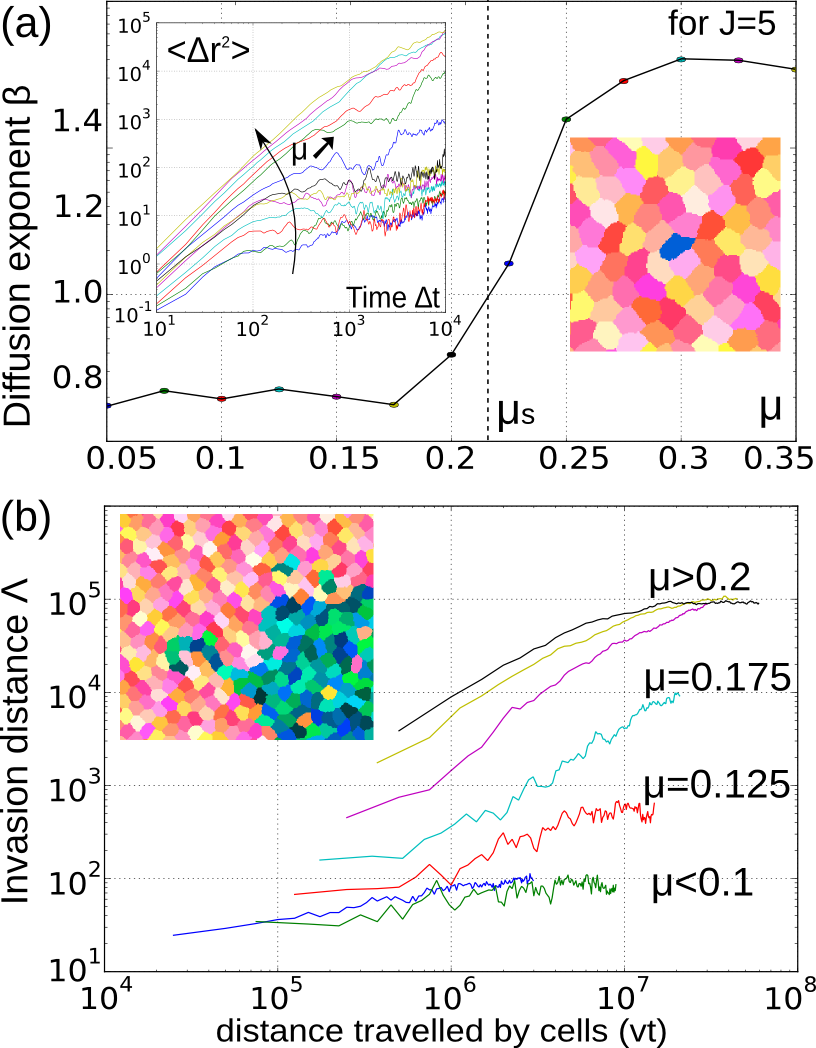}
\end{center}
\caption{(a) Diffusion exponent $\beta$ of a single motile cell in a tissue a non-motile cells. Left inset: Mean square displacement of single motile cells. Right inset: sample image of a single motile cell (blue) in the tissue. (b) Graph of the total invasion distance $\Lambda$ as a function of distance travelled by the motile cells (scaled time), for various motile forces ($J=5$).\label{invasion}}

\end{figure}

The minimal motile force $\mu_s$ required for the migration of a single cell in the passive tissue is first determined (see Video S5 to S7 \cite{SupMat}). 
Figure \ref{invasion}a shows the evolution of the mean square displacement of individual motile cells in a non-motile tissue. $\mu_s\approx 0.225$ corresponds to the cross-over between a sub-diffusive behaviour (cell trapped in the tissue) and a migratory behaviour. This threshold is here larger than the motility $\mu_c$ required for a motile population to undergo an EMT.

A tumour is now considered, modelled as a disc of 400 tumour cells (motile force $\mu>0$) surrounded by thousands of non-motile tissue cells. 
The invasive behaviour of the tumour is quantified by a scalar quantity $\Lambda(\mu, t)$ corresponding to the sum, for all the cells which have left the tumour, of their radial distance from the initial tumour boundary; it captures both the distance travelled and the number of cells migrating away. The time evolution of this quantity for a range of motile strengths is reported in figure  \ref{invasion}b.
For low cancer cell motility ($\mu<\mu_c$), cells are not migrating and no invasion occurs. For high motility ($\mu>\mu_s$), cells are motile enough to escape individually, leading to a very efficient invasion
. Cells eventually fully disperse in the tissue, and $\Lambda$ reaches then a plateau due to the finite size of the system. 
Although individual cells cannot invade when $\mu<\mu_s$, a significant amount of invasion however takes place in the range  $\mu_c<\mu<\mu_s$. In this regime, groups of cells, forming transient streams of 5 to 10 cells, are frequently seen leaving the tumour together, forming 5 to 10 cell long trains (see inset in figure \ref{invasion}b and Video S8 \& S9
cite{SupMat}). This behaviour is explained by the fact that coordinated motile cells generate together larger forces and can penetrate the surrounding tissue; it is therefore analogous to the ``tug of war'' process observed in vitro by Trepat et al. \cite{Trepat2009}. Collective invasion is favoured in this regime because single cell invasion is mechanically impossible, but small groups of cells can nevertheless coordinate and join forces. 

Although real tumour invasions are far more complex than the system studied here, the typical patterns of collective invasion only require straight-forward conditions: (i) an ability to coordinate cell migration (through the mechanisms suggested above, or otherwise) and (ii) a mechanical environment (surrounding tissue, or extra-cellular matrix) that can resist single cell invasion but fails if the force is slightly higher. This might explain why a particular cell type could exhibit different invasion strategies depending on the properties of its direct environment (represented here by $\mu_s$), as often observed \cite{Friedl2009b}. This also suggests that having a specialized cell at the tip of the invading group \cite{Gaggioli2007, Friedl2009b} is not necessary for coordination, but might be mechanically critical. In the common case of fibroblast led invasion, collective invasion might for instance proceed by i) a track being created in the matrix/surrounding tissue by the competent fibroblast cells \cite{Gaggioli2007} and ii) cancer cells streaming in the new space, strongly coordinated due to confinement in the track.

\subsection{Dynamic sorting}

The populations of cells considered in this paper all have the same membrane tension $J$, in order to avoid additional surface tension effects leading to cell sorting. Indeed, depending on the relative affinity of the different cell populations, mixing might be enhanced or prevented, interfering with the qualitative behaviour studied in this paper. In fact, invasion and sorting are strongly related phenomena: a poor ability to invade is analogous to a strong tendency to segregate.

\begin{figure}[!h]
\begin{center}
\includegraphics[width=3.4in]{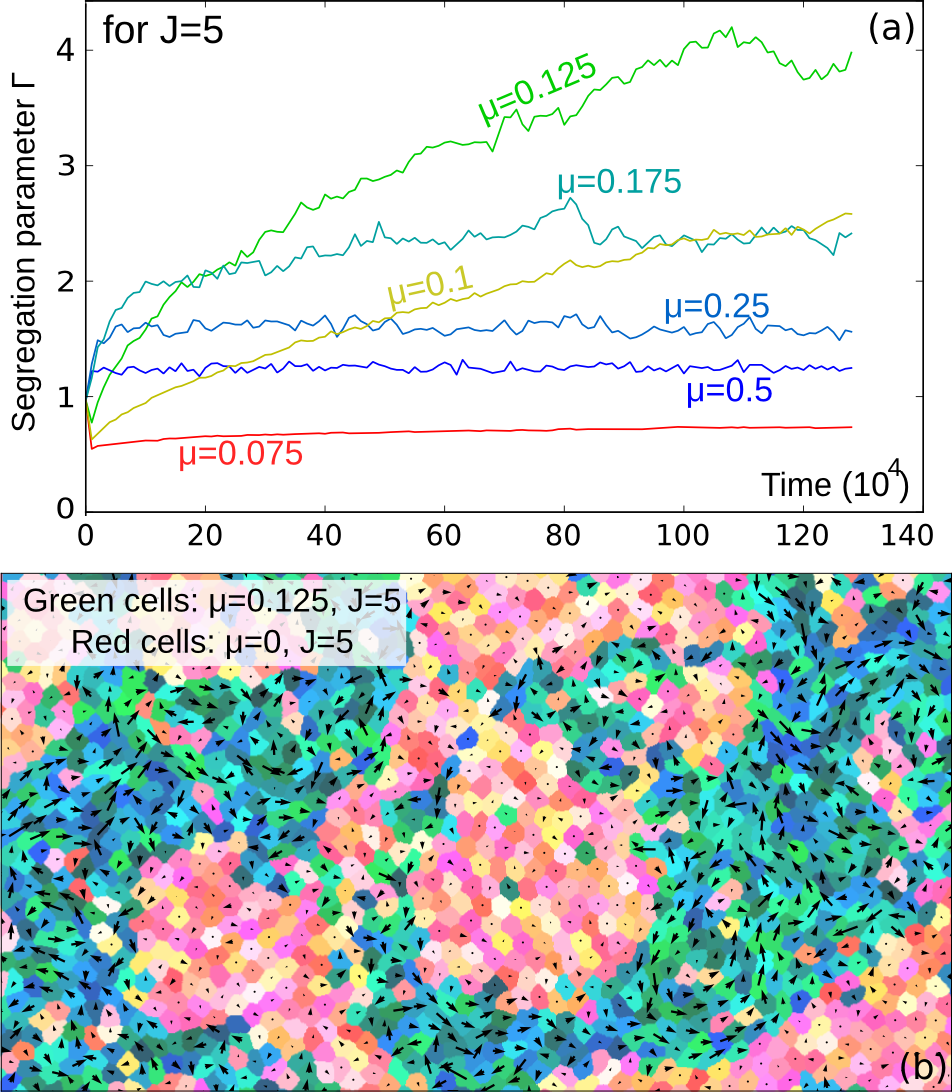}
\end{center}
\caption{(a) Evolution of $\Gamma(t)$ (the number of contacts between two motile cells divided by the number of contacts between motile and non motile cells) for different motile forces $\mu$ ($J=5$). (b) Segregating tissue at $t=10^6$ MCS with $\mu=0.125$ and $J=5$.\label{segreg}}
\end{figure}

To explore further this analogy, cell sorting is now studied in a system containing an equal proportion of motile and non-motile cells, initially arranged in a chessboard pattern. Cell sorting is monitored by calculating $\Gamma(t, \mu) = n_{MM} / n_{MN}$, where $n_{MM}$ is the number of contacts between motile cells, and $n_{NM}$ is the number of contacts between motile and non-motile cells. A large value of $\Gamma$ indicates a strong sorting effect, whereas a random distribution would lead to $\Gamma=1$. The figure \ref{segreg}(a) shows the evolution of $\Gamma$ as a function of time, for different values of the motile force $\mu$. For $\mu<\mu_c$, no sorting occurs since cells are not moving (static epithelium state). However, for motile forces just above $\mu_c$, sorting occurs, eventually leading to the formation of massive clusters of non-motile cells, surrounded by large streams of motile cells (figure \ref{segreg}(b) and video in supplementary material \cite{SupMat}). As the motile force further increases, sorting becomes nevertheless less and less efficient, as motile cells are becoming invasive and easily migrate through the groups of non-motile cells.

Earlier theoretical work on cell sorting has focused on adhesion differences to explain the separation of cell populations. Although cell motility is known to influence the dynamics of adhesion based segregation \cite{Belmonte2008}, its possible role as a driving force remained unexplored. The results presented here demonstrate that differences in motility can be sufficient to drive the sorting of cell populations, even without any significant difference in adhesive properties. A distinctive signature of such a process is the formation of streams of motile cells, and islands of non-motile groups. This prediction remains to be supported by experimental data.

\subsection{Discussion}

\subsubsection{A unified picture for collective migration}
The practical importance of the ideas introduced here relies most of all on their ability to provide, with parsimony, a unified picture of the patterns of collective migration. This work demonstrates that a non specific, mechanical coupling and a persistence time in the polarization direction of the cells are enough to reproduce a wide array of cellular behaviours commonly observed in vitro and in vivo. It reproduces the main features of the in vitro behaviour of endothelial cells away from a wound, and explains the mechanisms of large scale coordination when a wound is present. It shows that sheet migration is a robust feature of small groups of migrating cells, and predicts that the length scale required for sheet migration is related with the dynamics of the population in an unconfined condition. Finally, it provides guidance to interpret cell invasion processes and how these depend on matrix properties.

\subsubsection{Dynamic transitions}
The dynamics of a cell population has been primarily presented as a function of the motile force, for the sake of simplicity. However, the critical values $\mu_c$ and $\mu_s$ at which the qualitative transitions occur do depend on the other physical and biological parameters involved in this model. For instance, $\mu_c$ increases with $J$ (see figure \ref{diffusion}) and decreases with the persistence time $\tau$. As a result, transitions can be triggered by a variation of adhesion, cortical tension or response time as much as by a change in motile force $\mu$. The concept of dynamical transition, though common in physical systems, has a number of additional implications in a biological context. The results presented here show that minute variations of cellular and environmental parameters can trigger transitions between well defined phenotypes at the population scale (epithelium/sheet migration/mesenchyme, coordinated motion/disordered motion, no/collective/individual invasion). This might explain why such changes in phenotype are often difficult to associate with a proper lineage transition (or a change in gene expression) and how they depend on external parameters.

\subsubsection{Biomolecular signaling}

Highly organized behaviours are emerging without the need for specific cues of inter-cellular communications; this raises a number of questions concerning the role of known signalling pathways identified in association with individual or collective cell migration \cite{Affolter2005}. Cell-cell signalling might be seen either as a complementary (or reinforcing) mechanism, or as a way to compensate for collective effects. While any conclusions at this stage would be speculative, one might already comment on a few well established facts. Malignant fibroblasts have a reduced contact inhibition of locomotion (CIL) compared to non-malignant fibroblasts \cite{Abercrombie}. One might question to begin with why normal fibroblasts would need contact inhibition of locomotion, and a possible answer might lie in the phenomena of dynamic segregation. As demonstrated by figure \ref{segreg}, unless specific mechanisms are in place, motile fibroblasts would have a spontaneous tendency to stream and segregate rather than disperse in the tissue. CIL might therefore be a necessary behaviour to avoid segregation. This could also explain why changes in CIL can influence the emergence of collective effects in cell populations \cite{CarmonaFontaine2008}, and in particular help tumour cell invasion. If this model suggests generic mechanisms for collective behaviours, it also shifts the biological questions toward the understanding of how populations of active cells can accurately control their dynamic state in critical parts of development and tissue homeostasis. A number of current signalling mechanisms might have emerged from such evolutionary constraints.

\section{Materials and methods}

\subsection{Numerical model}

These simulations are implemented using the Cellular Potts Model (CPM) \cite{Graner1992}. The CPM is a lattice model: the state of the tissue is discrete, represented by an image of the system where the value $\sigma(k)$ of each pixel location $k$ codes for the identity of the cell covering that location. Cells can a priori adopt any shape on the lattice. The dynamics is introduced by minimizing an energy function of this state and other physical and biological parameters of the system. Adhesion and cortical tension are accounted for by a single parameter $J$ that provides an energy cost per unit of membrane length (in 2D) between cells. The cell volume is constrained to a reference value $v_0$, with a compressibility $\kappa^{-1}$. The energy function is minimized by randomly choosing one pixel at a time and testing if the energy can be lowered by transferring that pixel to a neighbouring cell. Fluctuations are introduced as part of this process, with a typical scale represented by a parameter $T$. Time is expressed in Monte Carlo Steps (MCS), where 1 MCS corresponds to an average of one iteration per pixel over the whole lattice.

\subsection{Motile force} Each cell generates a motile force along its polarization direction $\vect{n}_i$, with a strength $\mu_i$. These forces are here forces between each cell and its substrate. Interactions between cells being already accounted for by the membrane tension term. The energy function used in the CPM accounts for motility by adding, for each cell $i$, a time dependent energy gradient along $\vect{n}_i$. These essentially act as sources of energy that can drive cell motility. The function can be written:
\begin{equation}
E = \Sigma_{i} \frac{1}{2} \kappa (v_i - v_0)^2 + \Sigma_{k,l} J \cdot \left( 1 - \delta_{\sigma(k), \sigma(l)} \right) - \Sigma_i  \mu_i \vect{n}_i \cdot \vect{r}_i 
\end{equation}
where sums over $i$ are sums over all cells, and pairs (k,l) represent neighbouring pixels. $\delta_{\sigma(k), \sigma(l)} $ is 1 when the two pixels belong to the same cell, and 0 otherwise. The first term enforces the volume constraint, the second the membrane tension, and the third drives the motility. 

Dissipation in this model occurs due to the energy fluctuations, resulting in an effective viscous friction force between the cells and the substrate. The mean cell speed within the epithelium is mostly a linear function of $\mu/J$ for the range of motile forces used here. This is due to the fact that $J$ controls the amplitude of the membrane fluctuations, which in turn controls the dissipation between the cell and the substrate (fluctuation-dissipation theorem). In such a Potts model, the dynamics is therefore implicitly overdamped.

\subsection{Dynamics of the cell polarity}
The direction $\vect{n}$ of the motile force is determined by a feedback from its earlier displacements (unless specified otherwise, for instance concerning leader cells). $\vect{n}$ is oriented along the mean velocity of the cell over its past $\tau$ time-steps:
\begin{equation}
\vect{n}_i(t) =\frac{\left< \vect{v}_i\right>_{[t-\tau,t]}}{|\left< \vect{v}_i\right>_{[t-\tau,t]}|} 
\end{equation}
The parameter $\tau$ characterizes the time-scale at which cell polarity evolves and integrates changes in the mechanical environment, such as contact with boundaries or with other cells.

\subsection{Parameters used}
The simulations presented here have been performed using next-nearest neighbours interactions in the energy calculation, a preferred cell area ($v_0$) of 400 pixels (20x20), an energy fluctuation scale T=2.5, a compressibility $\kappa^{-1}$=0.5 and a memory time $\tau$=10 MCS. The results of figure \ref{leaders} have also been reproduced using next-next nearest neighbour interactions, using up to 6400 cells. The initial state of the system corresponds to a regular tessellation of the substrate, with each cell starting with a random orientation of its polarization. 

\subsection{Spatial correlations of the velocity field}
The analytical expression of $g(\vect{\delta r})$ is:
\begin{equation}
g(\vect{\delta r}) = \frac{ \left< \vect{v}\left(\vect{r}\right) \cdot \vect{v}\left(\vect{r} + \vect{R}\left(\vect{r}\right) \vect{\delta r}\right) \right>_{\vect{r}}}{ \left<v^2\right>_{\vect{r}}} 
\end{equation}
where $\vect{R}(\vect{r})$ is the transformation rotating $ \vect{v}\left(\vect{r}\right) $ along the horizontal $x$ axis.

\subsection*{Acknowledgements}
This work was supported by Engineering and Physical Sciences Research Council, EP/F058586/1. The author would like to warmly thank R. Adams' group, P. Hersen, L. Mahadevan, E. Sahai and B. Sanson for their valuable comments on this work.


\begin{thebibliography}{10}
\bibitem{Butler2009}
Butler LC et al. (2009) Cell shape changes indicate a role for extrinsic tensile forces in Drosophila germ-band extension. \textit{Nature Cell Biology} 11(7):859-64. 
\bibitem{Keller2000}
Keller R et al.
(2000) Mechanisms of convergence and extension by cell intercalation. \textit{Phil. Trans. R. Soc. Lond. B} 355:897-922. 
\bibitem{Concha1998}
Concha ML and Adams RJ (1998) Oriented cell divisions and cellular morphogenesis in the zebrafish gastrula and neurula: a time-lapse analysis. \textit{Development} 125:983-994.

\bibitem{CarmonaFontaine2008}
Carmona-Fontaine C et al. (2008) Contact inhibition of locomotion in vivo controls neural crest directional migration. \textit{Nature} 456(7224):957-61.

\bibitem{Lecaudey2008} 
Lecaudey V, Cakan-Akdogan G, Norton W H and Gilmour D (2008) Development 135(16):2695-705.

\bibitem{Friedl2009b}
Friedl P and Wolf K (2010) Plasticity of cell migration: a multiscale tuning model. \textit{J of Cell Bio} 188:11-19.

\bibitem{Friedl2009a}
Friedl P and Gilmour D (2009) {Collective cell migration in morphogenesis, regeneration and cancer}. \textit{Nature Reviews MCB} 10:445-457.



\bibitem{Weijer2009}
Weijer CJ (2009) {Collective cell migration in development}. \textit{J. Cell Sci.} 122:3215-3223.

\bibitem{Deisboeck2009}
Deisboeck TS and Couzin ID (2009)  {Collective behaviour in cancer cell populations}. \textit{BioEssays} 31:190-197.



\bibitem{Couzin2003}
Couzin ID and Krause J (2003) Self-Organization and Collective Behavior in Vertebrates. \textit{Advance in the Study of Behavior} 32:1-75.

\bibitem{Zhang2010}
Zhang HP, Be’er A, Florin EL and Swinney HL (2010) Collective motion and density fluctuations
in bacterial colonies. \textit{Proc. of Nat. Acad. Sci.} 107:3626–13630.
\bibitem{Schaller2010}
Schaller V, Weber C, Semmrich C, Frey E and Bausch A (2010) Polar patterns of driven filaments. \textit{Nature} 467:73.

\bibitem{Vicsek1995}
Vicsek T, Czirok A, Ben-Jacob E, Cohen I and Shochet O (1995) {Novel type of Phase Transition in a system of Self-Driven Particles}. \textit{Phys. Rev. Lett.} 75:1226.
\bibitem{Gregoire2004}
Gr\'egoire G and Chat\'e H(2004) {Onset of collective and cohesive motion}. \textit{Phys. Rev. Lett.} 92:025702.

\bibitem{Szabo2006}
Szabo B et al. (2006) Phase transition in the collective migration of tissue cells: Experiments and model. Phys Rev E 74:061908.

\bibitem{Szabo2010}
Szabo A et al. (2010) Collective cell motion in endothelial monolayers. \textit{Phys. Biol.} 7:046007.

\bibitem{Belmonte2008}
Belmonte JM, Thomas GL, Brunnet LG, de~Almeida RMC and Chat\'e H (2008) {Self-Propelled Particle Model for Cell-Sorting Phenomena}. \textit{Phys. Rev. Lett.} 100:248702.


\bibitem{Haga2005}
Haga H, Irahara C, Kobayashi R, Nakagaki T and Kawabata K (2005) Collective Movement of Epithelial Cells on a Collagen Gel Substrate. \textit{Biophys. J.} 88:2250-2256.

\bibitem{Poujade2007}
Poujade M et al. (2007) {Collective migration of an epithelial monolayer in response to a model wound}. \textit{Proc. of Nat. Acad. Sci.} 104:15988-15993.

\bibitem{Vitorino2008}
Vitorino P and Meyer T (2008) {Modular control of endothelial sheet migration}. \textit{Genes \& Development} 22:3268-3281.

\bibitem{Angelini2010}
Angelini TE, Hannezo E, Trepat X, Fredberg JJ and Weitz DA (2010) Cell Migration Driven by Cooperative Substrate Deformation Patterns. \textit{Phys Rev Lett} 104:168104.
\bibitem{Angelini2011}
Angelini TE et al. (2011) Glass-like dynamics of collective cell migration. \textit{Proc Natl Acad Sci USA} 108:4714.


\bibitem{Krieg2008}
Krieg M et al. (2008) Tensile forces govern germ-layer organization in zebrafish. \textit{Nature Cell Biology} 10:429-436.


\bibitem{Graner1992}
Graner F and Glazier JA (1992) {Simulation of Biological Cell Sorting Using a Two-Dimensional Extended Potts Model}.
\textit{Phys. Rev. Lett.} 69:2013-2016.

\bibitem{VanHaastert2010}
Van Haastert PJM (2010) {\em Chemotaxis: insights from the extending pseudopod}. \textit{J. of Cell Sci.} 123:3031-3037.
\bibitem{McCann2010}
McCann CP, Kriebel PW, Parent CA and Losert W (2010) {Cell speed, persistence and information transmission during signal relay and collective migration}.\textit{ J. of Cell Sci.} 123:1724-1731.

\bibitem{Peruani2007}
Peruani F and Morelli LG (2007) Self-Propelled Particles with Fluctuating Speed and Direction of Motion in Two Dimensions.
\textit{Phys. Rev. Lett.} 99:010602.



\bibitem{Tarin2005}
Tarin D (2005) The Fallacy of Epithelial Mesenchymal Transition in Neoplasia. \textit{Cancer Res.} 65:5996.

\bibitem{Thiery2006}
Thiery JP and Sleeman JP (2006) {\em Complex networks orchestrate epithelial-mesenchymal transitions}.
\textit{Nature MCB} 7:131.

\bibitem{Couzin2005}
Couzin ID, Krause J, Franks NR and Levin SA (2005) {\em Effective leadership and decision-making in animal groups on the move}. \textit{Nature} 433:513-518.

\bibitem{Trepat2009}
Trepat X et al. (2009) {Physical forces during collective cell migration}. \textit{Nature Physics} 5:426-430.

\bibitem{Gaggioli2007}
Gaggioli C et al. (2007) {Fibroblast-led collective invasion of carcinoma cells with differing roles for RhoGTPases in leading and following cells}. \textit{Nature Cell Biol.} 9:1392-1400.

\bibitem{Affolter2005}
Affolter M. and Weijer C.J. (2005) Signaling to Cytoskeletal
Dynamics during Chemotaxis. \textit{Developmental Cell} 9:19-34.

\bibitem{Abercrombie}
Abercrombie M (1979) Contact inhibition and malignancy. \textit{Nature} 281:259.

\bibitem{SupMat}
Supplementary Videos can be found at:

http://kalab.emma.cam.ac.uk/public/collmig/SM/
\end{thebibliography}
\end{document}